\documentclass[aps,twocolumn,prl,amssymb]{revtex4}
\usepackage{amsmath}
\usepackage{graphicx}
\usepackage{color}
\usepackage{setspace}
\usepackage{multirow}
\usepackage[]{titlesec}
\usepackage{enumerate}
\usepackage{natbib}
\definecolor{Gray}{gray}{0.9}
\definecolor{LightCyan}{rgb}{0.88,1,1}

\definecolor{Gray}{gray}{0.85}
\def\logTc{${\mathrm{log}}({\mathrm T}_c)$}
\def\ICSD{{\small ICSD}}

\def\ncalcs{20,000}

\begin{document}
\title{\Large Materials Cartography: Representing and Mining Material Space Using Structural and Electronic Fingerprints}

\author{Olexandr Isayev}
\affiliation{Laboratory for Molecular Modeling, Division of Chemical Biology and Medicinal Chemistry, UNC Eshelman School of Pharmacy, University of North Carolina, Chapel Hill, NC, 27599, USA.}
 \author{Denis Fourches}
\affiliation{Laboratory for Molecular Modeling, Division of Chemical Biology and Medicinal Chemistry, UNC Eshelman School of Pharmacy, University of North Carolina, Chapel Hill, NC, 27599, USA.}
\author{Eugene N. Muratov}
\affiliation{Laboratory for Molecular Modeling, Division of Chemical Biology and Medicinal Chemistry, UNC Eshelman School of Pharmacy, University of North Carolina, Chapel Hill, NC, 27599, USA.}
\author{Corey Oses}
\affiliation{Center for Materials Genomics, Duke University, Durham, NC 27708, USA}
\author{Kevin Rasch}
\affiliation{Center for Materials Genomics, Duke University, Durham, NC 27708, USA}
\author{Alexander Tropsha}
\email{alex_tropsha@unc.edu}
\affiliation{Laboratory for Molecular Modeling, Division of Chemical Biology and Medicinal Chemistry, UNC Eshelman School of Pharmacy, University of North Carolina, Chapel Hill, NC, 27599, USA.}
\author{Stefano Curtarolo}
\email[]{stefano@duke.edu}
\affiliation{Materials Science, Electrical Engineering, Physics and Chemistry, Duke University, Durham NC, 27708}
\affiliation{Center for Materials Genomics, Duke University, Durham, NC 27708, USA}

\date{\today}

\begin{abstract}
As the proliferation of high-throughput approaches in materials science is increasing the
wealth of data in the field, the gap between accumulated-information and derived-knowledge widens.
We address the issue of scientific discovery in materials databases by introducing novel analytical approaches
based on
structural and electronic materials fingerprints.
The framework is employed
to {\it (i)} query large databases of materials using similarity concepts,
{\it (ii)} map the connectivity of the materials space (i.e., as a materials cartogram) for rapidly identifying regions with unique organizations/properties,
and {\it (iii)} develop predictive Quantitative Materials Structure-Property Relationships (QMSPR) models for guiding materials design.
In this study, we test these fingerprints by seeking target material properties.  As a quantitative example, we model the critical temperatures of known superconductors.
Our novel materials fingerprinting and materials cartography approaches contribute to the emerging field of materials informatics by enabling effective computational tools to analyze,
visualize, model, and design new materials.
\end{abstract}
\maketitle

\section{Introduction}
\label{Intro}
Designing materials with desired physical and chemical properties is recognized as an outstanding challenge in materials research. \cite{Rajan_materialstoday_2005,nmatHT,Potyrailo_ACSCombSci_2011}
Material properties directly depend on a large number of key variables, often making the property prediction complex.
These variables include constitutive elements, crystal forms, geometrical and electronic characteristics; among others.
The rapid growth of materials research has led to accumulation of vast amounts of data.
For example, the Inorganic Crystal Structure Database (\ICSD)  includes more than 160,000 entries. \cite{ICSD} 
Experimental data are also included in other databases, such as MatWeb \cite{MatWeb} and MatBase. \cite{Matbase}
In addition, there are several large databases such as {\small AFLOWLIB}, \cite{aflowBZ,aflowSCINT} Materials Project, \cite{APL_Mater_Jain2013}
and Harvard Clean Energy \cite{Hachmann_JPCL_2011,Hachmann_EES_2014} that contain thousands of unique materials and their theoretically calculated properties.
These properties include electronic structure profiles estimated with quantum mechanical methods.
The latter databases have great potential to serve as a source of novel functional materials.
Promising candidates from these databases may in turn be selected for experimental confirmation using rational design approaches. \cite{MGI}

The rapidly growing compendium of experimental and theoretical materials data offers a unique opportunity for scientific discovery in materials databases.
Specialized data mining and data visualizing methods are being developed within the nascent field of materials informatics. \cite{Rajan_materialstoday_2005,Suh_MST_2009,Olivares-Amaya_EES_2011,Potyrailo_ACSCombSci_2011,nmatHT,Schuett_PRB_2014,Seko_PRB_2014}
Similar approaches have been extensively used in cheminformatics with resounding success.
For example, in many cases, these approaches have served to help identify and design small organic molecules with desired biological activity and acceptable environmental/human-health safety profiles. \cite{Laggner_NCB_2012,Besnard_Nature_2012,Cherkasov_JMC_2013,Lusci_JCIM_2013}
Application of cheminformatics approaches to materials science would allow researchers to {\it (i)} define, visualize, and navigate through the materials space, {\it (ii)} analyze and model structural and electronic characteristics of materials with regard to a particular physical or chemical property, and {\it (iii)} employ predictive materials informatics models to forecast the experimental properties of
{\it de novo} designed or untested materials.
Thus, rational design approaches in materials science constitutes a rapidly growing field. \cite{Olivares-Amaya_EES_2011,Balachandran_PRSA_2011,Kong_JCIM_2012,Balachandran_ActaCristB_2012,Srinivasan_MAT_2013,Schuett_PRB_2014,Seko_PRB_2014,Broderick_APL_2014,Dey_CMS_2014}

Herein, we introduce a novel materials fingerprinting approach.
We combine this with graph theory, similarity searches, and machine learning algorithms.
This enables the unique characterization, comparison, visualization, and design of materials.
We introduce the concept and describe the development of materials fingerprints that encode materials' band structures, density of states (DOS), crystallographic, and constitutional information.
We employ materials fingerprints to visualize this territory via advancing the new concept of {\it materials cartography}.
We show this technology identifies clusters of specific groups of materials with similar properties.
Finally, we develop Quantitative Materials Structure-Property Relationships (QMSPR) models that rely on materials fingerprints.
We then employ these models to discover novel materials with desired properties that lurk within the materials databases.

\section{Methods}
\label{methods}

{\small AFLOWLIB} is a database of Density Functional Theory calculations managed by the software package {\small AFLOW}. \cite{aflowPAPER,aflowlibPAPER}
At the time of writing, the {\small AFLOWLIB} database includes the results of calculations characterizing over \ncalcs\, crystals; representing about a quarter of the contents of the \ICSD\/. \cite{ICSD}
Of the characterized systems, roughly half are metallic and half are insulating.
{\small AFLOW} leverages the VASP Package \cite{vasp_cms1996} to calculate the total energy of a given crystal structure with PAW pseudopotentials \cite{PAW} and PBE \cite{PBE} exchange-correlation functional.
The entries of the repositories have been described previously. \cite{aflowBZ,aflowlibPAPER,aflowAPI}

\subsection{Data set of superconducting materials}
We have compiled experimental data for superconductivity critical temperatures T$_c$ for more than 700 records from the Handbook of Superconductivity, \cite{Poole_Superconductivity_2000}
CRC Handbook of Chemistry and Physics, \cite{Lide_CRC_2004} as well as SuperCon Database. \cite{SuperCon}
As we have shown recently, \cite{Fourches_JCIM_2010} data curation is a necessary step for any Quantitative Materials Structure-Property Relationship (QSAR) modeling.
In the compiled dataset, several T$_c$ values have been measured under strained conditions, such as different pressures and magnetic fields.
We have only kept records taken under standard pressure and  with no external magnetic fields.
For materials with variations in reported T$_c$ values in excess of 4 K, original references were revisited and records have been discarded when no reliable information was available.
T$_c$ values with a variation of less than 3 K have been averaged.
Of the remaining 465 materials (T$_c$ range of 0.1-133 K), most records show a variability in T$_c$ of $\pm$1 K between different sources.
Such a level of variability would be extremely influential in materials with low T$_c$ (T$_c\!<\!1$ K) because we have used the decimal logarithm of experimentally measured critical temperature (\logTc) as our target property.

In an effort to appropriately capture information inherent to materials over the full range of T$_c$, we have constructed two datasets for the development of three models.
The {\bf continuous model} serves to predict T$_c$ and utilizes records excluding materials with T$_c$ values less than 2 K.
This dataset consists of 295 unique materials with a \logTc\ range of 0.30-2.12.
The {\bf classification model} serves to predict the position of T$_c$ (above/below) with respect to the threshold T$_{thr}$ (unbiasedly set to 20K as observed in Figure \ref{fig_bands}(e), see Results and Discussion section). 
It utilizes records incorporating the aforementioned excluded materials, as well as lanthanum cuprate (La$_2$CuO$_4$, \ICSD\ \#19003).
Lanthanum cuprate had been previously discarded for high variability (T$_c$ = 21-39 K), but now satisfies the classification criteria.
This dataset consists of 464 materials (29 for T$_c\!>\!{\mathrm T}_{thr}$ and 435 set T$_c\!\leq\!{\mathrm T}_{thr}$).
Finally, the {\bf structural model} serves to identify structural components that most influence T$_c$. It utilizes the same dataset as the continuous model.

\subsection{Materials fingerprints}
Following the central paradigms of structure-property relationships, we assume that {\it (i)} properties of materials are a direct function of their structure and {\it (ii)} materials with similar structures (as determined by constitutional, topological, spatial, and electronic characteristics) are likely to have similar physical-chemical properties.

\begin{figure*}[t!]
\includegraphics[width=0.90\textwidth]{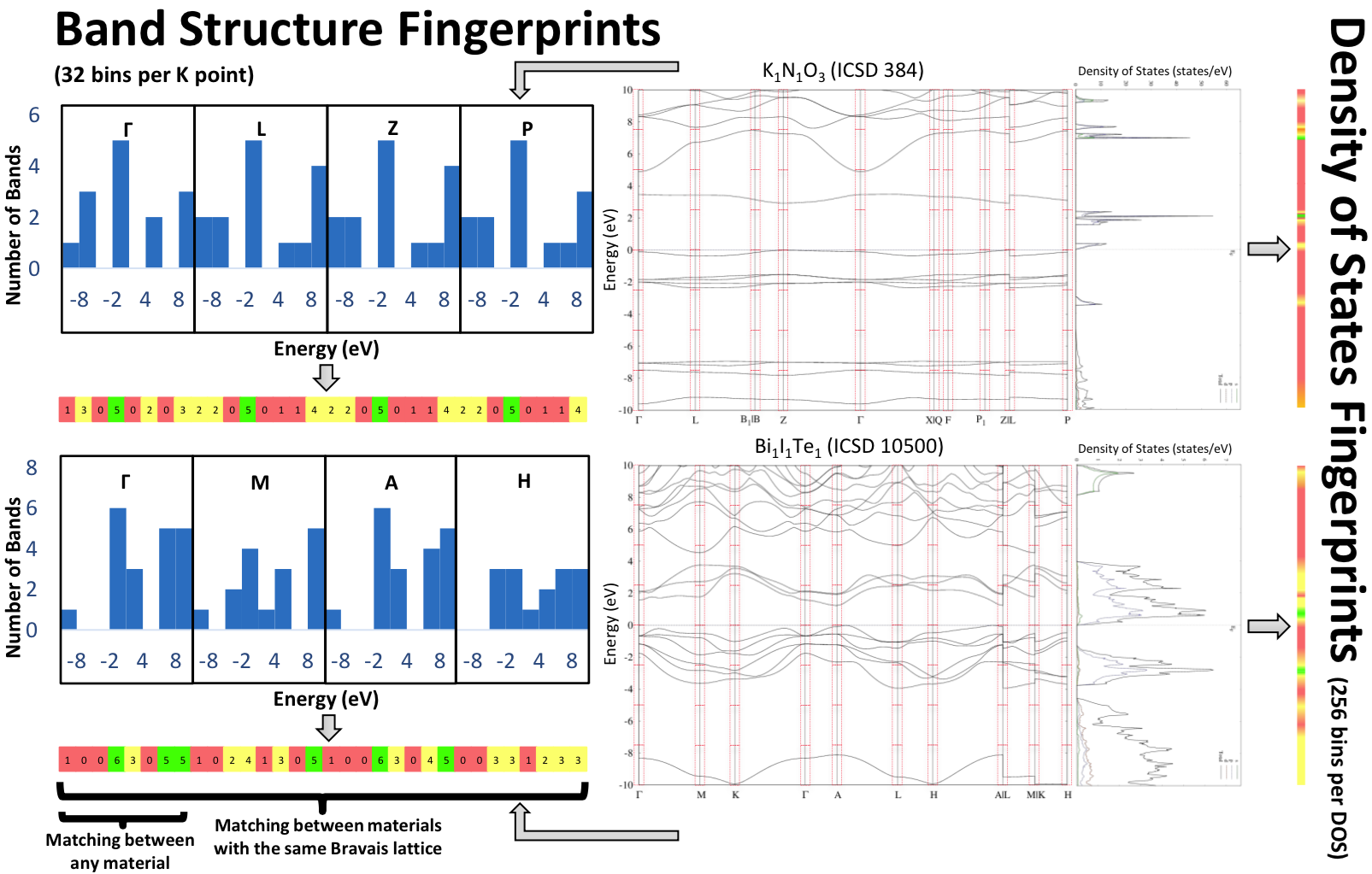}
\vspace{-3mm}
\caption{\small Construction of materials fingerprints from the band structure and the density of states. For simplicity we illustrate the idea of B-fingerprints with only 8 bins.}
\label{fig_fingerprints}
\end{figure*}

Thus, encoding material characteristics in the form of numerical arrays of descriptors, \cite{nmatHT,Schuett_PRB_2014} or {\it fingerprints}, \cite{Valle_ActaCristA_2010} enables the use of classical cheminformatics and machine-learning approaches to mine, visualize, and model any set of materials.
We have encoded the electronic structure diagram for each material as two distinct types of arrays (Figure \ref{fig_fingerprints}): {\it symmetry-dependent fingerprint} (band structure based “B-fingerprint”) and {\it symmetry-independent fingerprint} (density of state based “D-fingerprint”).

{\bf B-fingerprint.} Along every special high-symmetry point of the Brillouin zone (BZ), the energy diagram has been discretized into 32 bins to serve as our fingerprint array.
Each BZ has a unique set of high-symmetry points. \cite{aflowBZ}
The comparison set of high-symmetry points belonging to a single BZ type is considered {\it symmetry-dependent} like the B-fingerprint.
To name a few examples, the Brillouin zone path of a cubic lattice ($\Gamma–X–M–\Gamma–R–X\!\!\mid\!\! M–R$) are encoded with just four points ($\Gamma, M, R, X$), giving rise to a fingerprint array of length 128.
The body-centered orthorhombic lattice is more complex \cite{aflowBZ,aflowSCINT} ($\Gamma–X–L–T–W–R–X_1–Z–\Gamma–Y–S–W \!\!\!\mid\!\!\! L_1–Y\!\!\!\mid\!\!\! Y_1–Z$) and is represented by 13 points ($\Gamma, L, L_1, L_2, R, S, T, W, X, X_1, Y, Y_1, Z)$, giving a fingerprint array of length 416.
Conversely, the comparison of identical {\bf k}-points not specifically belonging to any BZ is always possible when only restricted to $\Gamma$.
Consequently, in the present work we limit our models to $\Gamma$ point B-fingerprint.

{\bf D-fingerprint.} A similar approach can be taken for the DOS diagrams, which is sampled in 256 bins (from min to max) and the magnitude of each bin is discretized in 32 bits.
Therefore, the D-fingerprint is a total of 1024 bytes.
Due to the complexity and limitations of the {\it symmetry-dependent} B-fingerprints, we have only generated {\it symmetry-independent} D-fingerprints.
The length of fingerprints is tunable depending on the objects, application, and other factors.
We have carefully designed the domain space and length of these fingerprints to avoid the issues of enhancing boundary effects or discarding important features.

{\bf SiRMS descriptors for materials.}
To characterize the structure of materials from several different perspectives, we have developed descriptors that can reflect their compositional, topological, and spatial (stereochemical) characteristics by utilizing structural descriptors similar to those used for small organic molecules.
Classical cheminformatics tools can only handle small organic molecules.
Therefore, we have modified the Simplex (SiRMS) approach \cite{Kuzmin_JCAMD_2008} based on our experience with mixtures \cite{Muratov_SC_2013,Muratov_MI_2012} in order to make this method suitable for computing descriptors for materials.

The SiRMS approach \cite{Kuzmin_JCAMD_2008} characterizes small organic molecules by splitting them into multiple molecular fragments called simplexes.
Simplexes are tetratomic fragments of fixed composition (1D), topology (2D), and chirality and symmetry (3D).
The occurrences of each of these fragments in a given compound are then counted.
As a result, each molecule of a given dataset can be characterized by its SiRMS fragment profiles.
These profiles take into account the atom types, connectivity, etc. \cite{Kuzmin_JCAMD_2008}
Here, we have adapted the SiRMS approach to describe materials with their fragmental compositions.

\begin{figure*}[t!]
 \includegraphics[width=0.90\textwidth]{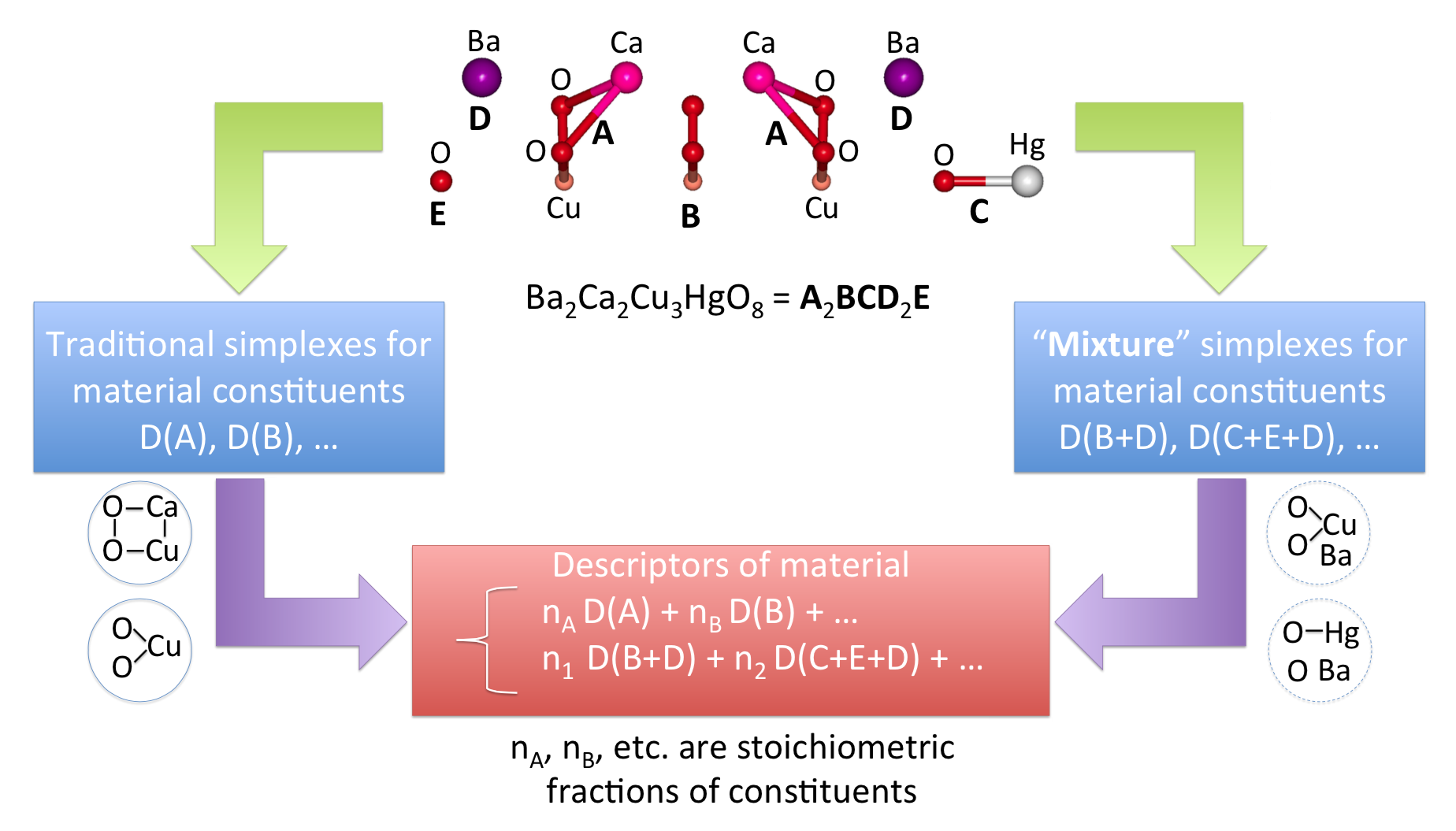}
 \vspace{-3mm}
 \caption{\small The generation of SiRMS descriptors for materials.}
 \label{fig_generation}
\end{figure*}

Every material is represented according to the structure of its crystal unit cell (Figure \ref{fig_generation}).
Computing SiRMS descriptors for materials is equivalent to the computation of SiRMS fragments for non-bonded molecular mixtures.
Bounded simplexes describe only a single component of the mixture.
Unbounded simplexes could either belong to a single component, or could span up to four components of the unit cell.
A special label is used during descriptor generation to distinguish ``mixture'' (belonging to different molecular moieties) simplexes from those incorporating elements from a single compound. \cite{Muratov_MI_2012}

Thus, the structure of every material is characterized by both bounded and unbounded SiRMS descriptors as illustrated in Figure \ref{fig_generation}.
The descriptor value of a given simplex fragment is equal to the number of its occurrences in the system.
In the case of materials, this value has been summed throughout all the constituents of a system; taking into account their stoichiometric ratios and crystal lattice (see Figure \ref{fig_generation}).
``Mixture'' descriptors are weighted according to the smallest stoichiometric ratio of constituents within this mixture, and added throughout all the mixtures in a system.
The atoms in simplexes are differentiated according to their type (element) and partial charge.
For the latter, the atoms are divided into six groups corresponding to their partial charge: $A\!\leq\!-2\!<\!B\!\leq\!-1\!<\!C\!\leq\!0\!<\!D\!\leq\!1\!<\!E\!\leq\!2\!<\!F$.
In addition, we have developed the special differentiation of atoms in simplexes to account for their groups on the periodic table.
That is, all elements belonging to the same group are encoded by the same symbol.

\subsection{Network Representation (Material Cartograms)}
To represent the library of materials as a network, we considered each material, encoded by its fingerprint, as a node.
Edges exist between nodes with similarities above certain thresholds.
In this study, we use fingerprint-based Tanimoto similarity and a threshold $S=0.7$.
This network representation of materials is defined as the graph $G(V,E)$, where $V=\left\{\nu_1|\nu_2\in L\right\}$ and
$E\!=\!\left\{(\nu_1,\nu_2)\mid\mathrm{sim}(\nu_1,\nu_2)\geq T\right\}$.
Here, $L$ denotes a materials library, $\mathrm{sim}(\nu_1,\nu_2)$
denotes a similarity between materials $\nu_1$ and $\nu_2$, and $T$ denotes a similarity threshold.

To examine if the material networks are scale-free, we analyzed the degree distributions of the networks.
Networks are considered scale-free if the distribution of the vertex degrees of the nodes follows the power law: $p(x)=kx^{-\alpha}$ where $k$ is the normalization constant, and $\alpha$ is the exponent.
The material networks have been visualized using the Gephi package. \cite{Bastian_ICWSM_2009}
The ForceAtlas 2 algorithm, \cite{Jacomy_PLoS_2014} a type of force-directed layout algorithm, has been used for the graph layout.
A force-directed layout algorithm considers a force between any two nodes, and minimizes the ``energy'' of the system by moving the nodes and changing the forces between them.
The algorithm guarantees that the topological similarity among nodes determines their vicinity, leading to accurate and
visually-informative representations of the materials space.

\section{Results and Discussion}
\label{results}
\subsection{Similarity search in the materials space}

In the first phase of this study, the optimized geometries, symmetries, band structures, and densities of states available in the {\small AFLOWLIB} consortium databases were converted into fingerprints, or arrays of numbers.

We encoded the electronic structure diagram for each material as two distinct types of fingerprints (Figure \ref{fig_fingerprints}): Band structure {\it symmetry-dependent} fingerprints (B-fingerprints), and density of states {\it symmetry-independent} fingerprints (D-fingerprints).
The B-fingerprint is defined as a collated digitalized histogram of energy eigenvalues sampled at the high-symmetry reciprocal points with 32 bins.
The D-fingerprint is a string containing 256 4-byte real numbers, each characterizing the strength of the DOS in one of the 256 bins dividing the [-10, 10] eV interval.
More details are in the Methods section.

This unique, condensed representation of materials enabled the use of cheminformatics methods, such as similarity searches, to retrieve materials with similar properties but different compositions from the {\small AFLOWLIB} database.
As an added benefit, our similarity search could also quickly find duplicate records.
For example, we have identified several BaTiO$_3$ records with identical fingerprints (\ICSD\ \#15453, \#27970, \#6102, and \#27965 in the {\small AFLOWLIB} database).
Thus, fingerprint representation afforded rapid identification of duplicates, which is the standard first step in our cheminformatics data curation workflow. \cite{Fourches_JCIM_2010}
It is well known that standard Density Functional Theory (DFT) has severe limitations in the description of excited states, and needs to be substituted
with more advanced approaches to characterize semiconductors and insulators. \cite{Hedin_GW_1965,GW,HSE,Liechtenstein1995,Cococcioni_reviewLDAU_2014}
However, there is a general trend of DFT errors being comparable in similar classes of systems.
These errors may thus be considered ``systematic'', and are irrelevant when one seeks only similarities between  materials.

The first test case is Gallium Arsenide, GaAs (\ICSD\ \#41674), a very important material for electronics \cite{INSPEC_PGA_1986} in the {\small AFLOWLIB} database.
GaAs is taken as the reference material, and the remaining \ncalcs+ materials from the {\small AFLOWLIB} database as the virtual screening library.
The pairwise similarity between GaAs and any of the materials represented by our D-fingerprints is computed using the Tanimoto similarity coefficient ($S$). \cite{Maggiora_JMC_2014}
The top five materials (GaP, Si, SnP, GeAs, InTe) retrieved show very high similarity ($S\!>\!0.8$) to GaAs, and all five are known to be semiconductor materials. \cite{Lide_CRC_2004,Littlewood_CRSSMS_1983,Madelung_Semiconductors_2004}

In addition, we have searched the {\small AFLOWLIB} database for materials similar to barium titanate (BaTiO$_3$)
with the Perovskite structure (\ICSD\ \#15453) using B-fingerprints.
BaTiO$_3$ is widely used as a ferroelectric ceramic or piezoelectric. \cite{Bhalla_MRI_2000}
Out of the six most similar materials with $S>0.8$, five (BiOBr, SrZrO$_3$, BaZrO$_3$, KTaO$_3$ and KNbO$_3$) are well known for their optical properties. \cite{Rabe_Ferroelectrics_2010}
The remaining material, cubic YbSe (\ICSD\ \#33675), is largely unexplored.
One can therefore formulate a testable hypothesis suggesting that this material may be ferroelectric or piezoelectric.

We also investigated the challenging case of topological insulators.
They form a rare group of insulating materials with conducting surface-segregated states (or interfaces) \cite{nmatTI} arising from a combination of spin-orbit coupling and time-reversal symmetry. \cite{RevModPhys.82.3045} 
Although DFT calculations conducted for materials in {\small AFLOWLIB} do not incorporate spin-orbit coupling for the most part, \cite{nmatTI} various topological insulators showed exceptionally high band-structure similarities, a manifesto for B-fingerprints.
The two materials most similar to Sb$_2$Te$_3$ \cite{RevModPhys.82.3045} (based on B-fingerprints) with $S\!>\!0.9$ were Bi$_2$Te$_3$ \cite{Chen09science,zhang_PRL_2009} and Sb$_2$Te$_2$Se. \cite{Xu2010arxiv1007}
Five out of six materials most similar to Bi$_2$Te$_2$Se \cite{Xu2010arxiv1007,Arakane2010NC} are also known topological insulators: Bi$_2$Te$_2$S, Bi$_2$Te$_3$, Sb$_2$Te$_2$Se, GeBi$_2$Te$_4$ \cite{Xu2010arxiv1007}, and Sb$_2$Se$_2$Te. \cite{nmatTI,Zhang_Nat.Phys._2009}

These examples demonstrate proof of concept and illustrate the power of simple yet uncommon fingerprint-based similarity searches for rapid and effective identification of materials with similar properties in large databases.
They also illuminate the intricate link between the structure and properties of materials by demonstrating that similar materials (as defined by their fingerprint similarity) have similar properties (such as being ferroelectric or insulators).
This observation sets the stage for building and exploring QMSPR models; as discussed below.

\subsection{Visualizing and exploring the materials space}

\begin{figure*}[t!]
 \includegraphics[width=0.95\textwidth]{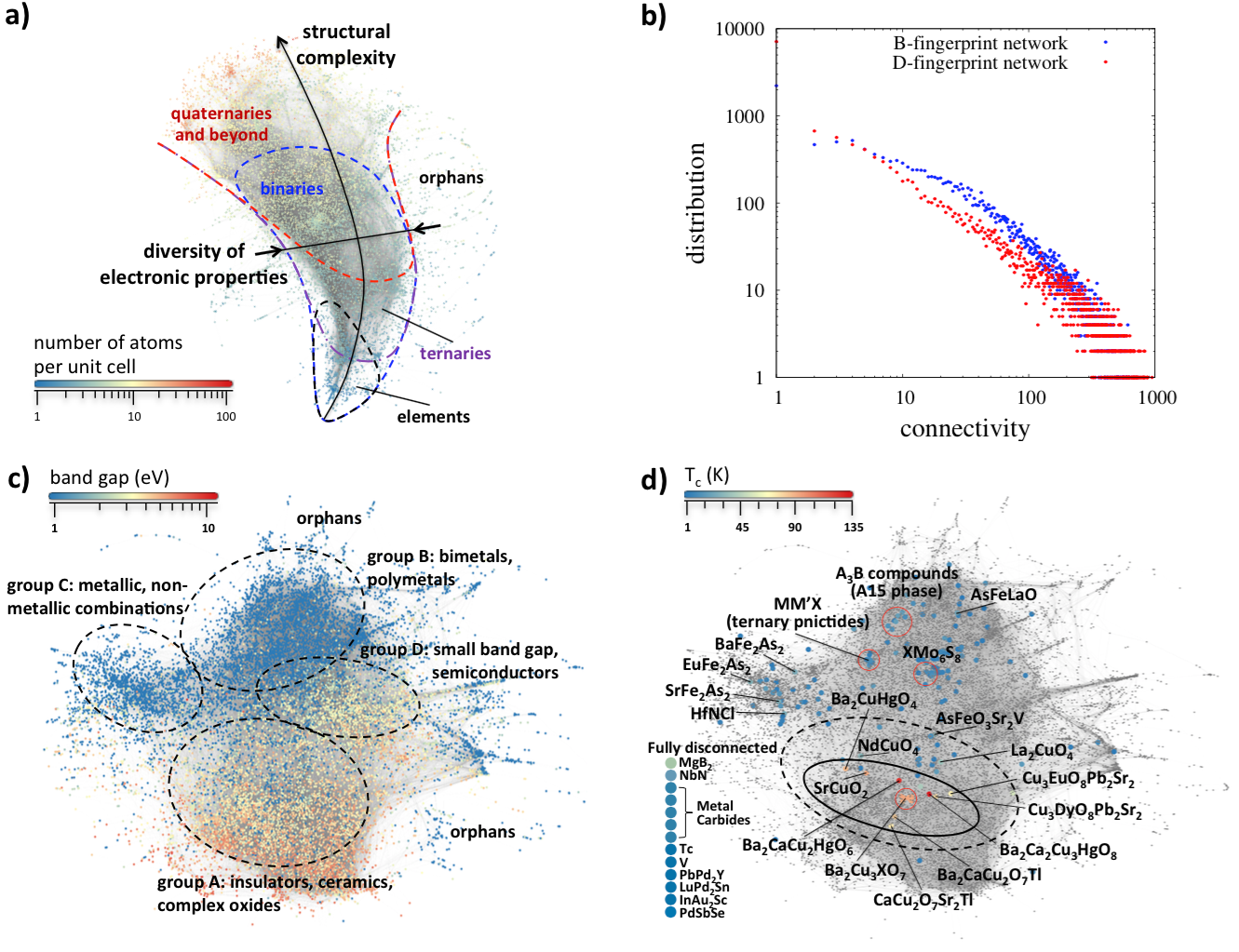}
 \vspace{-3mm}
 \caption{\small
  {\bf Materials cartograms (top) and B-fingerprint network representations (bottom).}
  {\bf a)} D-fingerprint network representation of materials.  Material are colored according to the number of atoms per unit cell.
  Regions corresponding to pure elements, binary, ternary and quaternary compounds are outlined.
  {\bf b)} Distribution of connectivity within the network.
  {\bf c)} Mapping band gaps of materials. Points colored in deep blue are metals; insulators are colored according to the band gap value. Four large communities are outlined.
  {\bf d)} Mapping the superconductivity critical temperature, ${\mathrm T}_c$, with relevant regions outlined.}
 \label{fig_cartograms}
 \label{fig_networks}
\end{figure*}

The use of fingerprint representation and similarity concepts led us to develop the materials network.
Compounds are mapped as nodes.
We use the {\it “force directed graph drawing’’} algorithm \cite{Herman_IEEEtvcg_2000} in which positions of the compound are initially taken randomly.
There is a force between the nodes: a repulsive Coulomb component and an optional attractive contribution with a spring constant equal to the Tanimoto coefficient between D-fingerprints (effective when $S\ge0.7$).
Two nodes are connected only when the coefficient is above the threshold.
The model is equilibrated through a series of heating and quenching steps.
Figure \ref{fig_cartograms}(a) shows the result in which we add Bezier-curved lines depicting regions of accumulation.
We shall refer to this approach to visualizing and analyzing materials and their properties as {\it materials cartography. }

The network shown in Figure \ref{fig_cartograms}(a) is colored according to overall complexity.
Pure systems, 79\% of the total 246 unary nodes, are confined in a small, enclosed region.
Binary nodes cover more configurational space, and 82\% of the 3700+ binaries lie in a compact region.
Ternaries are scattered. They mostly populate the center of the space (91\% of the 5300+ ternaries).
Quaternaries and beyond are located at the top part of the network (92\% of the 1080 nodes).
This region is the most distant from that of unary nodes, which tends to be disconnected from the others.
Indeed, overlap between binaries and ternaries is substantial.
The diversification of electronic properties and thickness of the compact envelope grows with structural complexity.
Orphans are defined as nodes with a very low degree of connectivity [only the vertices (materials) connected by edges are shown ($\sim$39\% of the database)].
Interestingly, of the 200 materials with connectivity smaller than 12, most are La-based (36 bimetallic and 126 polimetallic) or Ce-based (10 nodes).

The degree of connectivity is illustrated in Figure \ref{fig_cartograms}(b).
The panel indicates the log-log distribution of connectivity across the sample set.
The blue and red points measure the D-fingerprints (Figure \ref{fig_cartograms}(a)) and B-fingerprints connectivity (Figure \ref{fig_cartograms}(c)), respectively.
Table 1 in the Supporting Information contains relevant statistical information about the cartograms.
Although the power law distribution of Figure \ref{fig_cartograms}(b) is typical of scale-free networks and similar to many networks examined in cheminformatics and bioinformatics, \cite{Girvan_PNAS_2002,Newman_SiRev_2003,Yildirim_NB_2007} in our case, connectivity differs.
In previous examples, \cite{Girvan_PNAS_2002,Newman_SiRev_2003,Yildirim_NB_2007} most of the nodes have only a few connections; with a small minority being highly connected to a small set of ``hubs''. \cite{Jeong_Nature_2000,Barabasi_Science_1999}
In contrast, the {\small AFLOWLIB} network is highly heterogeneous:
most of the hubs' materials are concentrated along the long, narrow belt along the middle of the network.
The top 200 nodes (ranked by connectivity) are represented by 83 polymetallics (CoCrSi, Al$_2$Fe$_3$Si$_3$, Al$_8$Cr$_4$Y, etc.), 102 bimetallics (Al$_3$Mo, As$_3$W$_2$, FeZn$_{13}$, etc.), 14 common binary compounds (GeS, AsIn, etc.), and Boron (\ICSD\ \#165132).
This is not entirely surprising, since these materials are well studied and represent the lion's share of the \ICSD\  database.
Al$_3$FeSi$_2$ (\ICSD\ \#79710), an uncommonly used material, has the highest connectivity of 946.
Meanwhile, complex ceramics and exotic materials are relatively disconnected.

A second network, built with B-fingerprints, is illustrated in Figure \ref{fig_networks}(c).
While this network preserves most of the topological features described in the D-fingerprint case (Figure \ref{fig_cartograms}(a)), critical distinctions appear.
The B-fingerprint network separates metals from insulators.
Clustering and subsequent community analyses show four large groups of materials.
Group-A ($\sim$3000 materials) consists predominately of insulating compounds (63\%) and semiconductors (10\%).
Group-B distinctly consists of compounds with polymetallic character (70\% of $\sim$2500 materials).
In contrast, Group-C includes $\sim$500 zero band gap materials with non-metal atoms,
including halogenides, carbides, silicides, etc.
Lastly, Group-D has a mixed character with $\sim$300 small band gap (below 1.5 eV) materials; and $\sim$500 semimetals and semiconductors.

Lithium Scandium Diphosphate, LiScP$_2$O$_7$ (\ICSD\ \#91496), has the highest connectivity of 746 for the B-fingerprint network.
Very highly connected materials are nearly evenly distributed between groups A and B, forming dense clusters within their centers.
As in the case of the D-fingerprint network, the connectivity distribution follows a power law
(Figure \ref{fig_cartograms}(b), see Table 1 in the Supporting Information for additional statistics); indicating that this is a scale-free network.

To illustrate one possible application of the materials networks, we chose superconductivity---one of the most elusive challenges in solid-state physics.
We have compiled experimental data for 295 stoichiometric superconductors that were also available in {\small AFLOWLIB}.
All materials in the dataset are characterized with the fingerprints specified in the Methods section.
The dataset includes both prominently high temperature superconducting materials such as layered cuprates, ferropnictides, iron arsenides-122, MgB$_2$; as well as more conventional compounds such as A15, ternary pnictides, etc.
Our model does not consider the effect of phonons, which play a dominant role in many superconductors. \cite{tinkham_superconductivity}
High-throughput parameterization of phonon spectra is still in its infancy, \cite{curtarolo:Ru} and only recently have vibrational descriptors been
adapted to large databases. \cite{curtarolo:art96}
We envision that future development of vibrational fingerprints following these guidelines will capture similarities between
known, predicted, and verified superconductors (\textit{i.e.}, MgB$_2$ vs. LiB$_2$ \cite{curtarolo:art21,curtarolo:art26} and MgB$_2$ vs.
Fe-B compounds \cite{Kolmogorov_FeB_PRL2010,Gou_PRL_2013_FeB_superconductor}).

\begin{figure*}[t!]
 \includegraphics[width=0.99\textwidth]{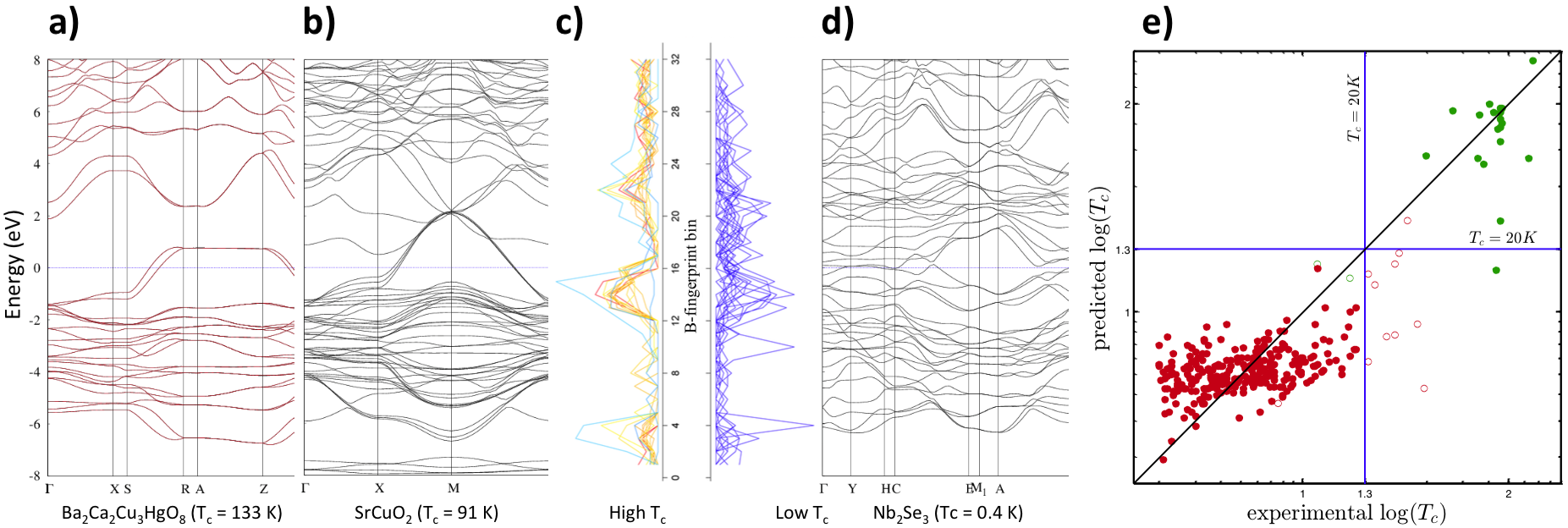}
 \vspace{-3mm}
 \caption{\small
  {\bf Comparison high-low T$_c$ aligned band structures and T$_c$ predictions}.
  {\bf a)} Band structure for Ba$_2$Ca$_2$Cu$_3$HgO$_8$, T$_c=$133 K.
  {\bf b)} Band structure for SrCuO$_2$, T$_c=$91 K. \cite{Takahashi_PSCC_1994_SrCuO2_superconductor}
  {\bf c)} Aligned B-fingerprints for the 15 materials with the highest and lowest T$_c$.
  {\bf d)} Band structure of Nb$_2$Se$_3$, T$_c=$0.4 K.
  {\bf e)} Plot of the predicted vs. experimental critical temperatures for the {\it continuous model}.
  Materials are color-coded according to the {\it classification model}: solid/open green (red) circles indicate
  correct/incorrect predictions in T$_c\!>\!{\mathrm T}_{thr}$ (T$_c\!\leq\!{\mathrm T}_{thr}$), respectively.}
 \label{fig_bands}
 \label{fig_Tc}
\end{figure*}

All materials are identified and marked on the B-fingerprints network, and are color-coded according to their critical temperature, T$_c$ (Figure \ref{fig_networks}(d)).
All high-T$_c$ superconductors are localized in a relatively compact region.
The distribution is centered on a tight group of Ba$_2$Cu$_3X$O$_7$ compounds (the so-called Y123, where $X$= Lanthanides).
The materials with the two highest T$_c$ values in our set are Ba$_2$Ca$_2$Cu$_3$HgO$_8$ (\ICSD\ \#75730, T$_c=$133 K) and Ba$_2$CaCu$_2$HgO$_6$ (\ICSD\ \#75725, T$_c=$125 K).
Their close grouping manifested a significant superconductivity hot-spot of materials with similar fingerprints.
We aligned the B-fingerprints for 15 superconductors with the highest T$_c$'s in Figure \ref{fig_bands}(c).

All of the top 15 high T$_c$ superconductors are layered cuprates, which have dominated high T$_c$ superconductor research since 1986. \cite{Bednorz_ZPBCM_1986}
These compounds are categorized as Charge-Transfer Mott Insulators (CTMI). \cite{Zaanen_PRL_1985}
There are three distinct bands that were conserved for these structures around -6, -1, and 4 eV relative to the Fermi energy at $\Gamma$ (within the simple DFT+$U$ description available in {\small AFLOWLIB}, Figure \ref{fig_bands}(c)).
These features are consistent with the three-band Hubbard-like picture characteristic of CTMIs. \cite{Manske_Superconductors_2004,Emery_PRL_1987}

Meanwhile, the fingerprint distribution for 15 materials with the lowest T$_c$ was random (Figure \ref{fig_bands}(c)).
The importance of band structure features in superconductivity has long been recognized. \cite{Zaanen_NPhys_2006,Micnas_RMP_1990,Orenstein_Science_2000}
Thus, the materials cartography based on the fingerprint network allows us to visualize this phenomenon concisely.

\subsection{Predictive QMSPR Modeling}

We developed QMSPR models (continuous, \cite{Bramer_PDM_2007} classification, and structural)
to compute superconducting properties of materials from their structural characteristics.
To achieve this objective, we compiled two superconductivity datasets consisting of
{\it (i)} 295 materials with continuous T$_c$ values ranging from 2 K to 133 K; and {\it (ii)} 464 materials with binary T$_c$ values.
The models were generated with Random Forest (RF) \cite{Breiman_ML_2001} and Partial Least Squares (PLS) \cite{Wold_CILS_2001} techniques.
These used both B- and D-fingerprints, as well as the Simplex (SiRMS) \cite{Kuzmin_JCAMD_2008} descriptors.
These fingerprints were adapted for materials modeling for the first time in this study (see Methods section).
Additionally, we incorporated atomic descriptors that differentiate by element, charge, and group within the periodic table.
Statistical characteristics for all 464 materials used for the QMSPR analysis
are reported in the Supporting Information (Tables 2-4).

Attempts to develop QMSPR models using B- and D-fingerprints for both datasets were not satisfactory, indicating that our fingerprints, while effective in qualitative clustering, do not contain enough information for quantitatively predicting target properties (QMSPR model acceptance criteria has been discussed previously \cite{Tropsha_MI_2010}).
Thus, we employed more sophisticated chemical fragment descriptors, such as SiRMS, \cite{Kuzmin_JCAMD_2008} and adapted them for materials modeling (see Methods section).

{\bf Continuous model.}
We constructed a {\it continuous model} which serves to predict the value of T$_c$ with a consensus RF- and PLS-SiRMS approach.
It has a cross-validation determination coefficient of $Q^2=0.66$ (five-fold external CV; see Table 2 of the Supporting Information).
Figure \ref{fig_Tc}(e) shows predicted versus experimental T$_c$ values for the continuous model: all materials having \logTc$\leq$1.3 were scattered, but within the correct range.
Interestingly, we notice that systems with \logTc$\geq$1.3 received higher accuracy, with the exceptions of
MgB$_{2}$ (\ICSD\ \#26675), Nb$_{3}$Ge (\ICSD\ \#26573), Cu$_{1}$Nd$_{2}$O$_{4}$ (\ICSD\ \#4203),
As$_{2}$Fe$_{2}$Sr (\ICSD\ \#163208), Ba$_{2}$CuHgO$_{4}$ (\ICSD\ \#75720), and ClHfN (\ICSD\ \#87795) (all highly underestimated).
Not surprisingly MgB$_2$ \cite{Buzea_SST_2001_MgB2} is an outlier in our statistics.  This is in agreement with the fact that to date
no superconductor with an electronic structure similar to MgB$_2$ has been found.

{\bf Classification model.}
By observing the existence of the threshold T$_{thr}$=20K (log(T$_{thr}$)=1.3), we developed a {\it classification model}.
It is based on the same RF-SiRMS technique, but it is strictly used to predict the position of T$_c$ with respect to the threshold, above or below.
The classification model has a balanced accuracy (BA) of 0.97 with five-fold external CV analysis.
The type of points in Figure \ref{fig_Tc}(e) illustrate the classification model outcome: solid/open green (red) circles for correct/incorrect
predictions in T$_c\!>\!{\mathrm T}_{thr}$ (T$_c\leq{\mathrm T}_{thr}$), respectively.

\begin{figure*}[htb!]
 \includegraphics[width=0.99\textwidth]{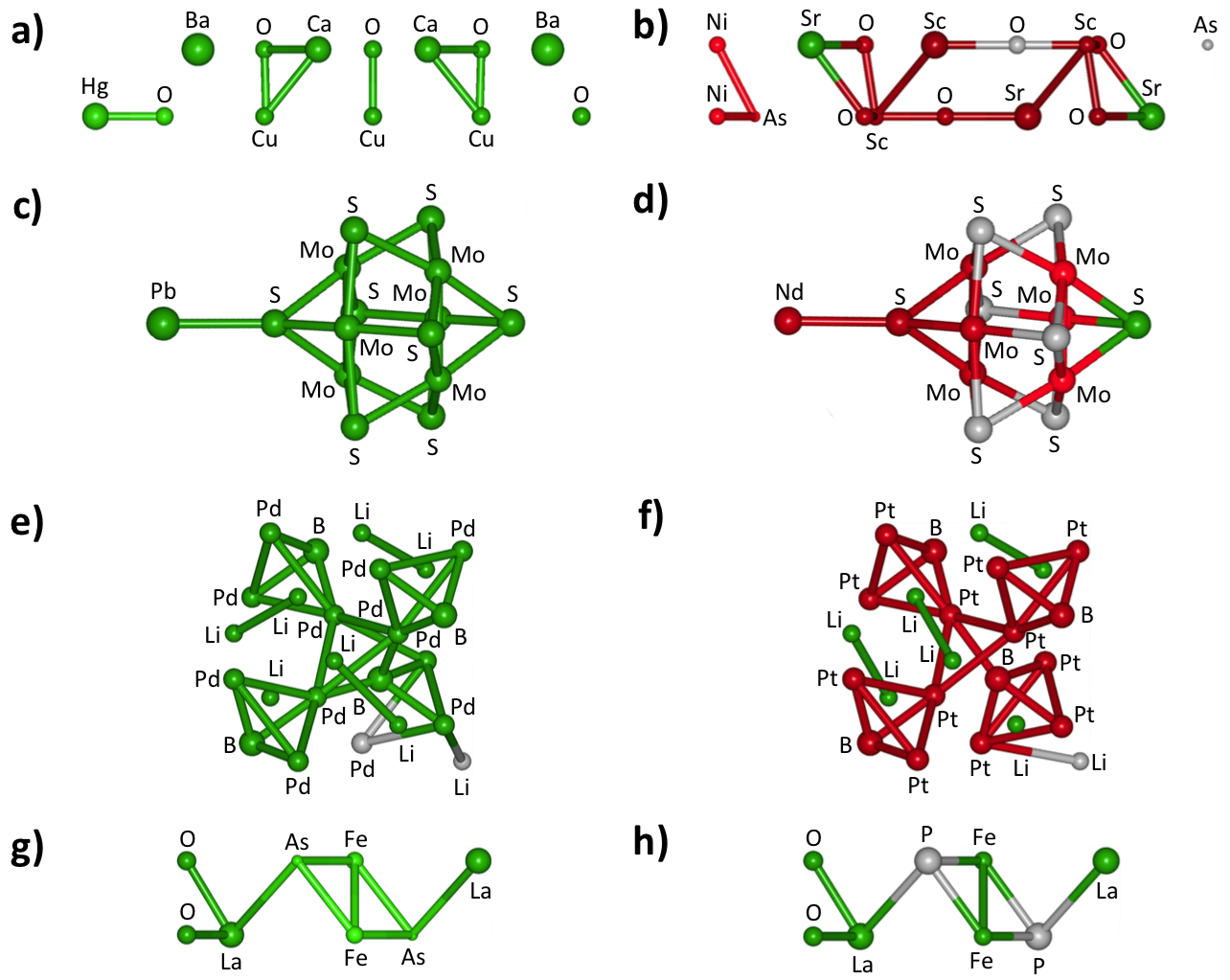}
 \vspace{-3mm}
 \caption{\small
  Materials color-coded according to atom contributions to \logTc.
  Atoms and structural fragments that decrease superconductivity critical temperatures
  are colored in red and those enhancing T$_c$ are shown in green.
  Uninfluential fragments are in gray.
  {\bf a)} Ba$_2$Ca$_2$Cu$_3$HgO$_8$ (\ICSD\ \#75730, \logTc=2.12);
  {\bf b)} As$_2$Ni$_2$O$_6$Sc$_2$Sr$_4$ (\ICSD\ \#180270, \logTc=0.44);
  {\bf c)} Mo$_6$PbS$_8$ (\ICSD\ \#644102, \logTc=1.13);
  {\bf d)} Mo$_6$NdS$_8$ (\ICSD\ \#603458, \logTc=0.54);
  {\bf e)} BLi$_2$Pd$_3$ (\ICSD\ \#84931, \logTc=0.89);
  {\bf f)} BLi$_2$Pt$_3$ (\ICSD\ \#84932, \logTc=0.49);
  {\bf g)} FeLaAsO (\ICSD\ \#163496, \logTc=1.32);
  {\bf h)} FeLaPO (\ICSD\ \#162724, \logTc=0.82). }
 \label{fig_fragments}
\end{figure*}

For T$_c\!\leq\!{\mathrm T}_{thr}$ and T$_c\!>\!{\mathrm T}_{thr}$, accuracies of prediction were 98\% and 90\% (cumulative 94\%).
(Figure \ref{fig_Tc}(e), see Table 3 in Supporting Information for additional statistics).
Among the 464 materials, ten systems with experimental T$_c\!>\!{\mathrm T}_{thr}$ were predicted to have
T$_c\!\leq\!{\mathrm T}_{thr}$)
[AsFeLaO (\ICSD\ \#163496), AsFeO$_{3}$Sr$_{2}$V (\ICSD\ \#165984), As$_{2}$EuFe$_{2}$ (\ICSD\ \#163210),
As$_{2}$Fe$_{2}$Sr (\ICSD\ \#163208), CuNd$_{2}$O$_{4}$ (\ICSD\ \#86754), As$_{2}$BaFe$_{2}$ (\ICSD\ \#166018), MgB$_{2}$, ClHfN, La$_{2}$CuO$_{4}$, and Nb$_{3}$Ge].
Only two with experimental T$_c\!\leq\!{\mathrm T}_{thr}$ were predicted with T$_c\!>\!{\mathrm T}_{thr}$ (AsFeLi (\ICSD\ \#168206), As$_{2}$CaFe$_{2}$ (\ICSD\ \#166016)).
Due to the spread around the threshold, additional information about borates and Fe-As compounds is required for proper training of the learning algorithm.

In the past, it has been shown that QSAR approaches can be used for the detection of mis-annotated chemical compounds,
a critical step in data curation. \cite{Fourches_JCIM_2010}
We have employed a similar approach here.
In our models, three materials, ReB$_2$ (\ICSD\ \#23871), Li$_2$Pd$_3$B (\ICSD\ \#84931), and La$_2$CuO$_4$ (\ICSD\ \#19003), were significantly mis-predicted.
More careful examination of the data revealed that T$_c$'s of ReB$_2$ and Li$_2$Pd$_3$B were incorrectly extracted from literature.
We also found that La$_2$CuO$_4$ had the largest variation of reported values within the dataset.
Therefore, it was excluded from the regression.
This approach illustrates that QMSPR modeling should be automatically implemented to reduce and correct erroneous entries.

{\bf Structural model.}
We also developed a {\it structural model} meant to capture the structural features that most influence T$_c$.
It employs SiRMS descriptors, PLS approaches, and five-fold external cross-validation.
The predictive performance of this model ($Q^2=0.61$) is comparable to that of the SiRMS-based RF model (see Table 2 in Supporting Information for additional statistics).
The top 10 statistically significant geometrical fragments and their contributions to T$_c$ variations are shown in Table 4 of the Supplementary Materials.
All descriptor contributions were converted to atomic contributions (details discussed previously \cite{Muratov_FMC_2010}) and related to material structures.
Examples of unit cell structures for pairs of similar materials with different T$_c$ values were color-coded according to
atomic contributions to T$_c$, and are shown in Figure \ref{fig_fragments}
(green for T$_c\!\uparrow$, red for T$_c\!\downarrow$, and gray for neutral).

Examples of fragments for materials having
T$_c\!>\!{\mathrm T}_{thr}$ [Ba$_2$Ca$_2$Cu$_3$HgO$_8$, \ICSD\ \#75730,
\logTc=2.12]
and T$_c\!\!\leq{\mathrm T}_{thr}$ [As$_2$Ni$_2$O$_6$Sc$_2$Sr$_4$, \ICSD\ \#180270, \logTc=0.44]
are shown in Figures \ref{fig_fragments}(a) and \ref{fig_fragments}(b), respectively.
They indicate that individual atom contributions are non-local as they strongly
depend upon the atomic environment (Figures \ref{fig_fragments}(c)-\ref{fig_fragments}(h)), {\it e.g.} Mo$_6$PbS$_8$ [\ICSD\ \#644102, \logTc=1.13] and Mo$_6$NdS$_8$ [\ICSD\ \#603458, \logTc=0.54]
differ by a substitution --- yet the difference in T$_c$ is substantial.
Furthermore, substitution of Nd for Pb affects contributions to the target property from all the remaining atoms in the unit cell
(Figure \ref{fig_fragments}(c) and \ref{fig_fragments}(d)).
The same observation holds for BLi$_2$Pd$_3$ [\ICSD\ \#84931, \logTc=0.89] and BLi$_2$Pt$_3$ [\ICSD\ \#84932, \logTc=0.49] Figure \ref{fig_fragments}(e) and
 \ref{fig_fragments}(f); as well as FeLaAsO [\ICSD\ \#163496, \logTc=1.32] and FeLaPO [\ICSD\ \#162724, \logTc=0.82] Figure \ref{fig_fragments}(g) and \ref{fig_fragments}(h).

\section{Conclusion}

With high-throughput approaches in materials science increasing the data-driven content of the field, the gap between accumulated-information and derived knowledge widens.
The issue can be overcome by adapting the data-analysis approaches developed during the last decade for chem- and bio-informatics.

Our study gives an example of this.
We introduce novel materials fingerprint descriptors that lead to the generation of networks called {\it “materials cartograms’’}: nodes represent compounds; connections represent similarities.
The representation can identify regions with distinct physical and chemical properties,
the key step in searching for interesting, yet unknown compounds.

Starting from atomic-composition, bond-topology, structure-geometry, and electronic properties of materials publicly available in the {\small AFLOWLIB} repository,
we have introduced cheminformatics models leveraging novel materials fingerprints.
Within our formalism, simple band-structure and DOS fingerprints are adequate to locate metals, semiconductors, topological insulators, piezoelectrics, and superconductors.
More complex QMSPR modeling \cite{Kuzmin_JCAMD_2008} are used to tackle qualitative and quantitative values of superconducting critical temperature and geometrical features helping/hindering criticality, including the use of SiRMS descriptors. \cite{Kuzmin_JCAMD_2008}

In summary, the fingerprinting cartography introduced in this work has demonstrated its utility in an initial set of problems.
This shows the possibility of designing new materials and gaining insight into the relationship between the structure
and physical properties of materials. Further advances in the analysis and exploration of databases may become the
foundation for rationally designing novel compounds with desired properties.

\section*{Acknowledgement}

We thank Drs. Marco Buongiorno Nardelli, Stefano Sanvito, Ohad Levy, Amir Natan, Gus Hart, Allison Stelling, Luis Agapito, and Cheng-Ing Chia
for various technical discussions that have contributed to the results reported in this article.
A.T. acknowledges support from DOD-ONR (N00014-13-1-0028), ITS Research Computing Center at UNC, and the
Russian Scientific Foundation (No. 14-43-00024) for partial support.
S.C. acknowledges support from DOD-ONR (N00014-13-1-0030, N00014-13-1-0635),
DOE (DE-AC02-05CH11231, specifically BES grant \# EDCBEE), and
the Duke University Center for Materials Genomics.
C.O. acknowledges support from the National Science Foundation Graduate Research Fellowship under Grant No. DGF1106401.
We also acknowledge the CRAY corporation for computational support.

\cleardoublepage
\cleardoublepage


\newcommand{\Ozolins}{Ozoli\c{n}\v{s}}

\end{document}